\documentclass[twocolumn,showpacs,preprintnumbers,amsmath,amssymb]{revtex4}

\usepackage{graphicx}
\usepackage{dcolumn}
\usepackage{bm}

\begin{document}

\title{Epidemic spread in weighted scale-free networks}
\author{YAN Gang$^{1}$}
\author{ZHOU Tao$^{1,2}$}
\author{WANG Jie$^{1}$}
\author{FU Zhong-Qian$^{1}$}
\author{WANG Bing-Hong$^{2}$}
\email{bhwang@ustc.edu.cn,Fax:+86-551-3603574}
\affiliation{%
$^{1}$Electronic Science and Technology,\\
University of Science and Technology of China,\\
Hefei Anhui, 230026, PR China \\
$^{2}$Nonlinear Science Center and Department of Modern Physics,\\
University of Science and Technology of China,\\
Hefei Anhui, 230026, PR China }%

\date{\today}

\begin{abstract}
In this letter, we investigate the detailed epidemic spreading
process in scale-free networks with links' weights that denote
familiarity between two individuals and find that spreading
velocity reaches a peak quickly then decays in a power-law form.
Numerical study exhibits that the nodes with larger strength is
preferential to be infected, but the hierarchical dynamics are not
clearly found, which is different from the well-known result in
unweighed network case. In addition, also by numerical study, we
demonstrate that larger dispersion of weight of networks results
in slower spreading, which indicates that epidemic spreads more
quickly on unweighted scale-free networks than on weighted
scale-free networks with the same condition.
\end{abstract}

\pacs{89.75.-k, 89.75.Hc, 87.23.Ge, 05.70.Ln}

\maketitle

Many social, biological, and communication systems can be properly
described as complex networks with vertices representing
individuals or organizations and links mimicking the interactions
among them. Recently, the ubiquity of a power-law degree
distribution in real-life networks has attracted a lot of
attention\cite{Reviews}. Examples of such networks (scale-free
networks or SF networks for short) are numerous: these include the
Internet, the World Wide Web, social networks of acquaintance or
other relations between individuals, metabolic networks, integer
networks, food webs, etc.\cite{Networks}. The ultimate goal of the
study on topological structure of networks is to understand and
explain the workings of systems built upon those networks, for
instance, to understand how the topology of the World Wide Web
affects Web surfing and search engines, how the structure of
social networks affects the spread of diseases, information,
rumors or other things, how the structure of a food web affects
population dynamics, and so on.

Recent studies on epidemic spreading in SF networks indicate a
particular relevance in the case of networks characterized by
complex topologies and very heterogeneous
structures\cite{Reviews,Stanley 01} that in many cases present us
with new epidemic propagation scenarios\cite{Ep,B.B.P.V 04}, such
as absence of any epidemic threshold\cite{Ep}, hierarchical spread
of epidemic outbreaks\cite{B.B.P.V 04}, and so on. The new
scenarios are of practical interest in computer virus diffusion
and the spreading of diseases in heterogeneous populations.
Further more, they also raise new questions on how to protect the
networks and find optimal strategies for the deployment of
immunization resources\cite{Imm}. However, so far, studies of
epidemic spread just focus on unweighted SF networks, and a
detailed inspection of epidemic spreading process in weighted SF
networks is still missing while real networks, such as population
and Internet, are obviously scale-free and with links' weights
that denote familiarity between two individuals(like people or
computers), respectively. One can easily take cognizance of how
the links' weights affect the epidemic spreading process. For
instance, if your little son gets flu, then you will be infected
in all probability, since you two contact each other very
frequently(i. e. of large familiarity). By contraries, it is
unlikely that you will be infected by your unfamiliar colleague
just because of saying hello to him this morning.

\begin{figure}
\scalebox{0.8}[0.8]{\includegraphics{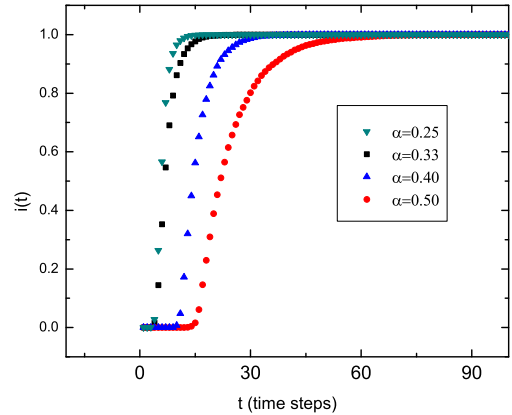}}
\caption{\label{fig:epsart}Density of infected individuals versus
time in a BBV network with $N=10^4,\delta=3.0,\omega_{{\rm
0}}=1.0$ and $m=3$, the four numerical curves $i(t)$ correspond
with parameter $\alpha =$0.5, 0.4, 0.33 and 0.25 respectively,
form bottom to top.}
\end{figure}

In this letter, we intend to provide a first analysis of the time
evolution of epidemic spreading in weighted SF networks. The
weighted SF network model used in this letter is one of the most
well-known model introduced by Barrat, Barthelemy, and Vespignani
(BBV networks)\cite{B.B.V 04}, whose degree, strength and weight
distributions are power-law distributions with heavy tails. The
BBV model suggests that two main ingredients of self-organization
of a network in a weighted scale-free structure are strength
preferential attachment and weights' dynamics. These point to the
facts that most networks continuously grow by the addition of new
vertices, new vertices are preferentially attached to existing
vertices with larger strength, and the creation of new links will
introduce variations of the existing weight distribution. More
precisely, the weight of each new edge is fixed to value
$\omega_{0}$; if a new vertex linked to an existing vertex $i$,
then the local rearrangement of weights between $i$ and its
neighbors $j$ according to the simple rule
\begin{equation}
\omega_{ij}\rightarrow \omega_{ij}+\Delta\omega_{ij}
\end{equation}
where
\begin{equation}
\Delta\omega_{ij}=\delta\frac{\omega_{ij}}{s_{i}}
\end{equation}
$s_{i}$ is the strength of node $i$, expressed by
$s_{i}=\sum_{j}\omega_{ij}$. This rule considers that the
establishment of a new edge of weight $\omega_{0}$ with the vertex
$i$ induces a total increase of traffic $\delta$ that is
proportionally distributed among the edges departing from the
vertex according to their weights. Since BBV networks are of the
same properties (e.g. power-law distribution of degree, strength
and weight) as many social networks (e.g. friendship networks and
scientists collaboration networks) and technical networks (e.g.
Internet and WWW), it is reasonable to investigate epidemic
spreading on BBV networks.

\begin{figure}
\scalebox{0.75}[0.8]{\includegraphics{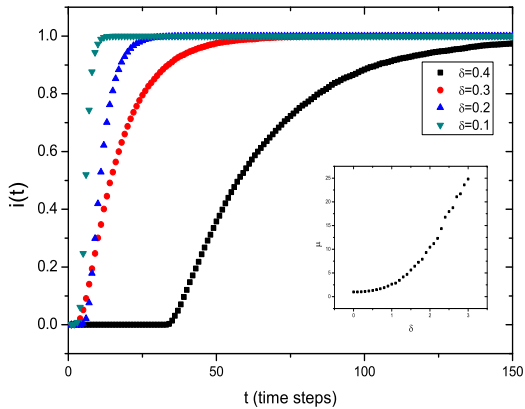}}
\caption{\label{fig:epsart} Density of infected individuals versus
time in a BBV network with $N=10^4,\alpha=2.0,\omega_{{\rm
0}}=1.0$ and $m=3$, the four numerical curves $i(t)$ correspond
with parameter $\delta=$0.4, 0.3, 0.2 and 0.1 respectively, form
bottom to top. The inset shows the relationship between the
dispersion of weight ($\mu$) and the value of $\delta$.}
\end{figure}

\begin{figure}
\scalebox{0.9}[0.8]{\includegraphics{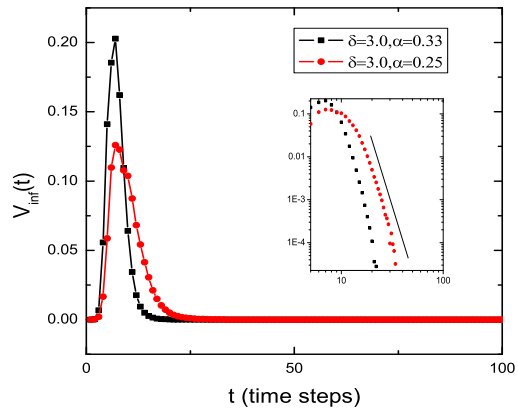}}
\caption{\label{fig:epsart} Spreading velocity at each time $t$ in
a BBV network with $N=10^4,\delta=3.0,\omega_{{\rm 0}}=1.0$ and
$m=3$,when $\alpha=0.33(square)$ and $\alpha=0.25(circle)$. The
inset shows the curves in log-log plot. The data are averaged over
200 experiments.}
\end{figure}

In order to study the dynamical evolution of epidemic spreading we
shall focus on the susceptible-infected (SI) model in which
individuals can be in two discrete states, either susceptible or
infected\cite{Aderson 92}. Each individual is represented by a
vertex of the network and the links are the connections between
individuals along which the infection may spread. The total
population(the size of the network) $N$ is assumed to be constant
thus if $S(t)$ and $I(t)$ are the number of susceptible and
infected individuals at time $t$, respectively, then $N=S(t)+I(t)$
. In weighted networks, we define the infection transmission by
the spreading rate,
\begin{equation}
\lambda_{ij}=(\frac{\omega_{ij}}{\omega_M})^\alpha, \alpha>0
\end{equation}
at which susceptible individual $i$ acquire the infection from the
infected neighbor $j$, where $\alpha$ is a positive constant and
$\omega_{M}$ is the largest value of $w_{ij}$ in the network.
Obviously, more familiar two individuals(i.e. with larger weight)
may infect each other with greater probability. According to
Eq.(3), one can quickly obtain the probability that an susceptible
individual $i$ will be infected at the present time step is:
\begin{equation}
\lambda_i(t)=1-\prod_{j\in N_i(t)}(1-\lambda_{ij})
\end{equation}
where $N_i(t)$ is the set of all $i$'s infected neighbors at time
$t$.

\begin{figure}
\scalebox{0.95}[0.8]{\includegraphics{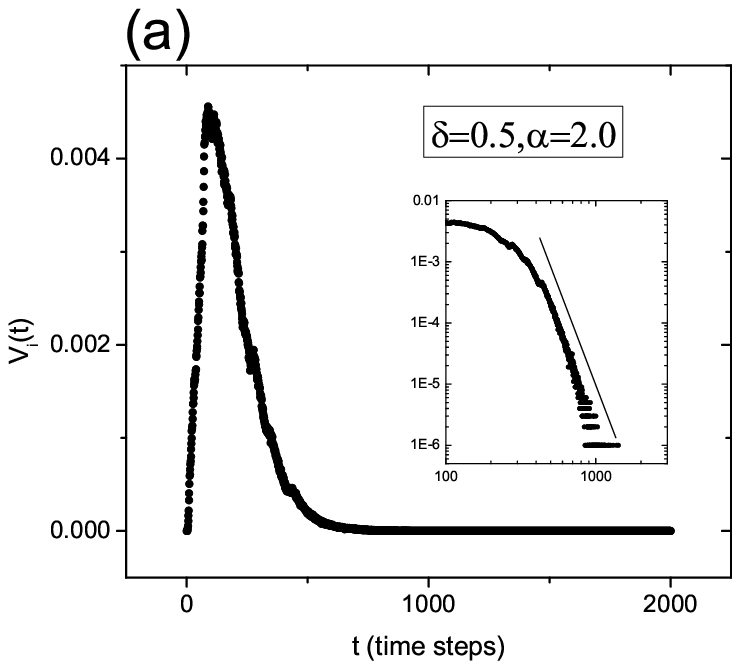}}
\scalebox{0.87}[0.8]{\includegraphics{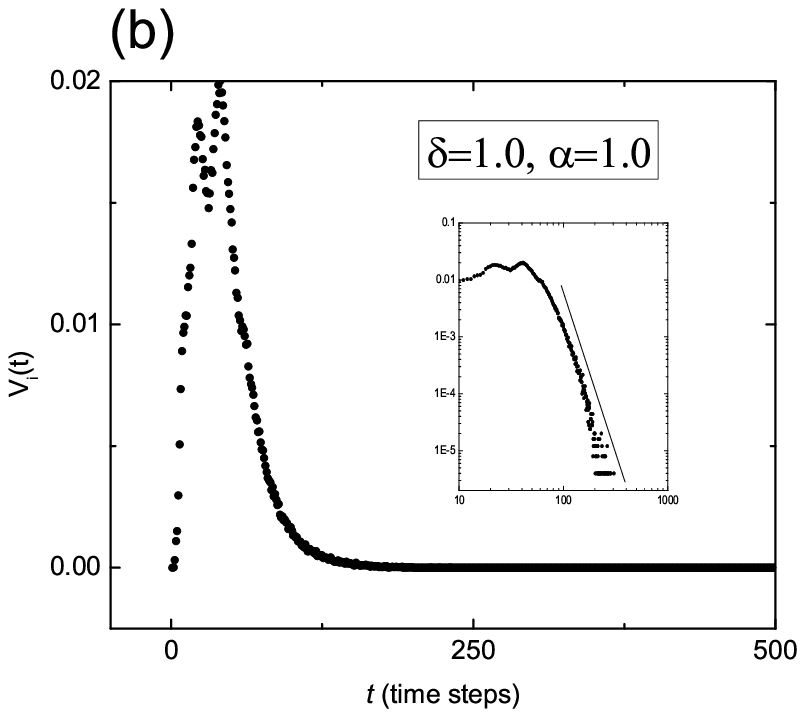}}
\caption{\label{fig:epsart} Consider the cases of $\delta<1.0$ and
$\delta=1.0$. (a) and (b) show the spreading velocity at each time
$t$ in a BBV network with $N=10^4,\omega_{0}=1.0$ and $m=3$, when
$\delta=0.5, \alpha=2.0$ and $\delta=1.0, \alpha=1.0$,
respectively. The inset shows the curves in log-log plot. The data
are averaged over 200 experiments.}
\end{figure}

We start by selecting one vertex randomly and assume it is
infected. The diseases or computer virus will spread in the
networks in according with the rule of Eq.(3). In \emph{Fig.1}, we
plot the density of infected individuals versus time in a BBV
network with $N=10^4,\delta=3.0,\omega_{{\rm 0}}=1.0$ and $m=3$.
Since $\frac{\omega_{ij}}{\omega_M}\leq1$, the smaller $\alpha$
is, the more quickly infection spreads. It is natural that larger
value of $\delta$ induces larger dispersion of weight of networks.
Then, a direct question is that how the value of $\delta$ impacts
epidemic spreading behavior. In \emph{Fig.2}, we show that
epidemic spreads more quickly while $\delta$ is smaller. In other
words, larger dispersion of weight of networks results in slower
spreading. That means epidemic spreads more quickly on unweighted
scale-free networks than on weighted scale-free networks with the
same condition.

Obviously, all the individuals will be infected in the limit of
long time as $\lim_{t\rightarrow \infty}i(t)=1$. For the sake of
finding optimal strategies to protect individuals from being
infected, we will study the details of spreading velocity at the
outbreak moment. The spreading velocity is defined as:
\begin{equation}
V_{{\rm inf}}(t)=\frac{di(t)}{dt}\approx \frac{I(t)-I(t-1)}{N}
\end{equation}
where $i(t)=\frac{I(t)}{N}$. We account the number of newly
infected vertices at each time step and report the spreading
velocity in {\it Fig.3}. Apparently, the spreading velocity goes
up to a peak quickly that similar to the unweighted network
cases\cite{B.B.P.V 04}, leaving us very short response time to
develop control measures. Moreover, what's new and interesting,
velocity decays following power-law form after the ``peak time".
At the moment of infection outbreaks, the number of infected
individuals is very small, as well as after a very long time from
the outbreak, the number of susceptible individuals is very small.
Thus when $t$ is very small(close to zero) or very large, the
spreading velocity is close to zero, one can see the corresponding
simulation result in {\it Fig.3}. One may think that the velocity
follows power-law behavior just because of the extreme case of
$\delta>1.0$. Now we consider the case of $\delta<1.0$ and
$\delta=1.0$. {\it Fig.4} shows spreading velocity at each time
$t$ in a BBV network with $N=10^4,\omega_{0}=1.0$ and $m=3$, when
$\delta=0.5, \alpha=2.0$ and $\delta=1.0, \alpha=1.0$,
respectively. It is obvious that epidemic spreading behavior dose
not show sensitive dependence on the parameter $\delta$, the
reason of that fact will be explored deeply in our future
publications.

In order to give a more precise characterization of the epidemic
diffusion through the weighted networks, we measure the average
strength of newly infected vertices at time $t$, define as:
\begin{equation}
{\bar{S}_{{\rm inf}}(t)}=\frac{\sum_{s}s[I_{{\rm s}}(t)-I_{{\rm
s}}(t-1)]}{I(t)-I(t-1)}
\end{equation}
where $I_{{\rm s}}(t)$ is the number of infected vertices with
strength $s$. {\it Fig. 4} shows the average strength of newly
infected vertices $\bar{S}_{{\rm inf}}(t)$ as a function of time
$t$, and the curves exhibit that $\bar{S}_{{\rm inf}}(t)$ displays
a power-law behavior for large $t$, $\bar{S}_{{\rm inf}}(t)\propto
t^{-\gamma}$, which is remarkably different from the clear
hierarchical feature on unweighted networks\cite{B.B.P.V 04}.

It is explicit that the individuals with larger strength are much
more dangerous when they are infected, rather than the ones with
smaller strength, thus if one want to protect most individuals
being infected, the susceptible individuals with larger strength
must be protect foremost. In {\it Fig. 5}, one can find that the
individuals with larger strength is preferential to be infected,
which means there is little time leaving us to find the ``{\it
Large Individuals}" and isolate them. Therefore, at the outbreak
moment of disease or computer virus, the dense crowd or pivotal
servers must be protected primarily. Of course, the outcome is not
a good news for practical operators, but it may be relevant for
the development of containment strategies.

\begin{figure}
\scalebox{0.75}[0.8]{\includegraphics{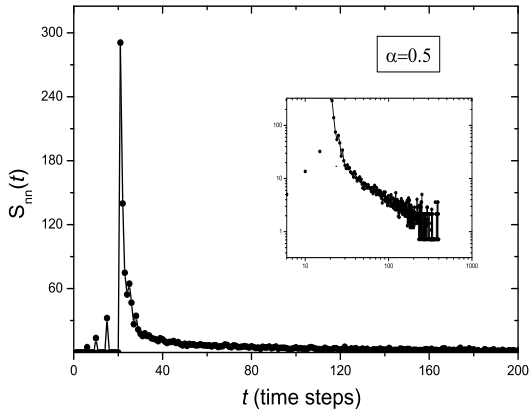}}
\caption{\label{fig:epsart} Behavior of average strength of the
newly infected vertices at time $t$ for SI model spreading in a
BBV network with $N=10^4,\delta=3.0,\omega_{{\rm 0}}=1.0$ and
$m=3$, the inset shows that $\bar{S}_{{\rm inf}}(t)$ represents
power-law behavior, $\bar{S}_{{\rm inf}}(t)\propto t^{-\gamma}$.}
\end{figure}

In summary, we have studied epidemic spreading process in BBV
networks, and the present results provide a clear picture of the
infection propagation in weighted SF networks. The numerical
studies show that spreading velocity $V_{{\rm inf}}(t)$ and
average strength of newly infected vertices $\bar{S}_{{\rm
inf}}(t)$ present power-law time behavior for large $t$, which is
remarkably different from infection propagation in unweighted
networks. Also by numerical study, we demonstrate that larger
dispersion of weight of networks results in slower spreading,
which indicates that epidemic spreads more quickly on unweighted
scale-free networks than on weighted scale-free networks with the
same condition. These results indicate that not only the
topological structures of networks but also the links' weights
affect the epidemic spreading process. Further more, the detailed
study of behavior of average strength of the newly infected
vertices may be relevant for the development of containment
strategies.

However, up to now, there are so many important and fundamental
problems that puzzle us and haven't been referred to in the
present letter. Some of them have been partially solved and will
be publicized in further publication, and others are still
unanswered. At the end of this letter, we will list part of them.
How to analyze the average density of infected individuals versus
time at the outbreak moment in weighted SF networks, and how about
the dynamic behavior after ``peak time"? Is the mean-field theory
appropriate to solve this problem? How to design a optimal
containment strategy, and how about the effective for various
strategies, such as to protect vertices at random, to protect
vertices purposefully, to cut off links at random, to cut off
links purposefully, and so on?

This work has been partially supported by the State Key
Development Programme of Basic Research (973 Project) of China,
the National Natural Science Foundation of China under Grant
No.70271070, 70471033 and 10472116, the Specialized Research Fund
for the Doctoral Program of Higher Education (SRFDP
No.20020358009), and the foundation for graduate students of
University of Science and Technology of China under Grant No.
USTC-SS-0501.

\end{document}